\documentclass[12pt]{article}
\usepackage{times}
\usepackage{geometry}
\usepackage[utf8]{inputenc}
\usepackage{enumitem,amssymb}
\usepackage{ragged2e}
\usepackage{graphicx}
\usepackage{epsfig}
\usepackage{color}
\usepackage{txfonts}
\usepackage{natbib,twoopt}
\newlist{thematic}{itemize}{8}
\setlist[thematic]{label=$\square$}
\usepackage{pifont}
\newcommand{\cmark}{\ding{51}}%
\newcommand{\done}{\rlap{$\square$}{\raisebox{2pt}{\large\hspace{1pt}\cmark}}%
\hspace{-2.5pt}}

\begin{document}
\raggedright
\huge
Astro2020 Science White Paper \linebreak

Mapping Galaxy Clusters in the Distant Universe\linebreak
\normalsize

\noindent \textbf{Thematic Areas:} \hspace*{60pt} $\square$ Planetary Systems \hspace*{10pt} $\square$ Star and Planet Formation \hspace*{20pt}\linebreak
$\square$ Formation and Evolution of Compact Objects \hspace*{31pt}  $\square$ Cosmology and Fundamental Physics \linebreak
  $\square$  Stars and Stellar Evolution \hspace*{1pt} $\square$ Resolved Stellar Populations and their Environments \hspace*{40pt} \linebreak
$\done$     Galaxy Evolution   \hspace*{45pt} $\square$             Multi-Messenger Astronomy and Astrophysics \hspace*{65pt} \linebreak

\textbf{Principal Authors:}

Name: Helmut Dannerbauer
 \linebreak						
Institution: Instituto de Astrof\'{i}sica de Canarias (IAC), E-38205 La Laguna, Tenerife, Spain
\linebreak
Universidad de La Laguna, Dpto. Astrof\'{i}sica, E-38206 La Laguna, Tenerife, Spain
 \linebreak
Email: helmut@iac.es
 \linebreak
Phone: +34 922 605 251
 \linebreak
 
Name: Eelco van Kampen
 \linebreak						
Institution: European Southern Observatory, Karl-Schwarzschild-Straße 2, 85748 Garching, Germany
 \linebreak
Email: evkampen@eso.org
 \linebreak
Phone: +49 89 3200 6875
 \linebreak

\textbf{Co-authors:} 
\linebreak
Jose Afonso (IA), Paola Andreani (ESO), Fabrizio Arrigoni Battaia (MPA), Frank Bertoldi (AIfA), Caitlin Casey (University of Texas), Chian-Chou Chen (ESO), David L. Clements (Imperial College), Carlos De Breuck (ESO), Brenda Frye (Steward Observatory), James Geach (Herts), Kevin Harrington (AIfA), Masao Hayashi (NAOJ), Shuowen Jin (IAC, ULL), Pamela Klaassen (STFC), Kotaro Kohno (University of Tokyo) Matthew D. Lehnert (IAP), Israel Matute (IA), Tony Mroczkowski (ESO), Allison Noble (MIT), Ciro Pappalardo (IA), Yoichi Tamura (Nagoya), Jorge Zavala (University of Texas)
\linebreak

\textbf{Abstract:}
\linebreak
\justify
\vspace{-0.85cm}
We present the science case for mapping several thousand galaxy (proto)clusters at $z=1-10$ with a large aperture single dish sub-mm facility, producing a high-redshift counterpart to local large surveys of rich clusters like the well-studied Abell catalogue.
Principal goals of a large survey of distant clusters are the evolution of galaxy clusters over cosmic time and the impact of environment on the evolution and formation of galaxies. To make a big leap forward in this emerging research field, the community would benefit from a large-format, wide-band, direct-detection spectrometer (e.g., based on MKID technology), covering a wide field of $\sim$1~square degree and a frequency coverage from 70 to 700~GHz. 

\pagebreak
\section{Introduction}
\begin{figure*}[!hb]
\begin{centering} 
\includegraphics[width=\linewidth]{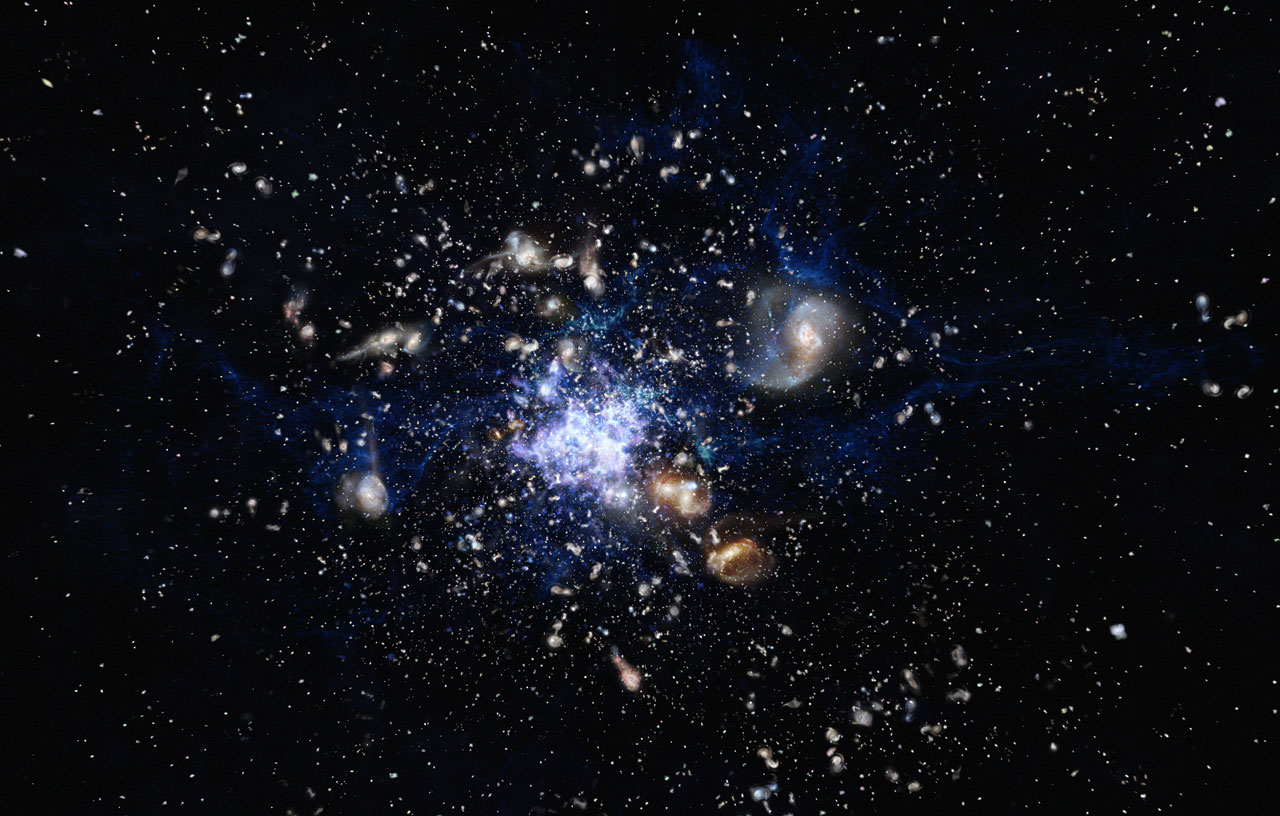}
\caption{This artist’s impression depicts the formation of a galaxy cluster in the early Universe. The galaxies are vigorously forming new stars and interacting with each other. They are observed as Far-Infrared/Sub-mm Galaxies or Dusty Star-Forming Galaxies, and are the focus of this white paper. Credit: ESO/M. Kornmesser. Courtesy from ESO Press Release October 2014 on \citet{dan14} 
}\label{fig:spiderweb}
\end{centering} 
\end{figure*}

\justify
As the first structures to collapse, galaxy clusters have to be seen as the earliest fingerprint of galaxy formation and evolution \citep[see][for a review]{kra12}. These structures grew hierarchically through the merging and accretion of smaller units of galaxy halos.  These are dominated by (very) young galaxies displaying intense bursts of star-formation --- the dusty star-forming galaxy population \citep[DSFGs; see][for a review]{cas14}. They are rich of molecular gas and heavily obscured by dust, thus prime targets for far-infrared/submm facilities. These galaxy systems are most probably the progenitors of elliptical galaxies \citep[e.g.,][]{lut01,ivi13} which dominate local galaxy clusters such as Virgo or Coma. Fig.~\ref{fig:spiderweb} --- based on work presented in \citet{dan14} on a protocluster at $z=2.2$ (age of the Universe: 3 Gyrs) --- shows how violent galaxy clusters could be in the distant Universe. These high density regions are  remarkable environments for investigating the physical processes responsible for the triggering and suppression of star formation and black hole activity. In this white paper, we will motivate the need of a systematic study of (proto)galaxy clusters with a large single dish sub-mm telescope at a high-site to make big leaps forward in this emerging research field. 

\section{Why we need a large single-dish sub-mm telescope}
\subsection{State-of-the-Art}
A future systematic (sub)mm survey of high-redshift (proto)galaxy clusters would address the following current fundamental drawbacks:
\newline\newline
\noindent \textbf{Small, heterogenous samples:} The number of known spectroscopically confirmed (proto)clusters beyond $z=2$ is still less than 50 \citep[see][for a review]{ove16}. The datasets collected up to now are still highly heterogeneous and likely with strong selection bias, making it impossible to obtain a complete picture of the build-up of galaxy clusters over cosmic time. Furthermore, the small samples of high-redshift (proto)clusters we have are often sparsely sampled. 
\newline
\newline
\noindent \textbf{Large spatial distribution on the sky:}
According to simulations and observations \citep[e.g.,][]{mul15,cas16}, high-redshift galaxy clusters and their infall region should cover a linear extension of up to 30 Mpc, which corresponds to about 30$^{\prime}$ at $z\sim1-7$. Therefore, studying and understanding the key epoch of cluster formation really needs to cover areas of up to ons equare degree, something which is all but impossible with adequate sensitivity with any current sub-mm facility.
\newline\newline
\noindent \textbf{Lack of cold ISM measurements of galaxy cluster members:}
The past decade has seen a rise in several hundred detections of the cold molecular gas supply that fuel the star formation in the distant Universe, albeit focusing mostly on isolated field galaxies \citep[e.g.,][]{car13,tac18}. However, the number of published cold gas measurements of galaxies located in galaxy clusters at $z>1$ is fairly low ($<50$). Especially thanks to ALMA, this number did increase significantly in the past two years \citep[e.g.,][]{coo18,dan17,hay17,hay18,nob17,nob19,rud17,sta17,tad19}, but remains low nonetheless. 
\newline\newline
\noindent \textbf{Survey speed:} 
Currently, ALMA allows for a survey speed of at most a few square arcminutes per hour for the brightest CO lines \citep{pop16} to yield a sufficient number of cluster galaxies with L$>$ L$_{\ast}$ at $z\approx1$. The survey speed for a given line depends on the sensitivity to that line and the primary beam at its frequency (which determines the number of pointings needed for a given map). Obtaining an area of a square degree with ALMA would take at least 1000 hours for the line with the highest survey speed, CO(5-4), which is hard to interpret physically. CO(3-2) would take around 6000 hours with ALMA, whereas CO(2-1) would need three times that \citep{pop16}, so one degree surveys for the lower transition lines are prohibitively expensive with ALMA. Another property of ALMA is the relatively narrow bandwidth (just below 8 GHz) which make spectral scans slow too.

\subsection{Proposed Study}
Our main goal is to map the formation and early evolution of galaxies and their environment, including (proto)clusters and other large-scale structures which are forming at the same time in the early Universe ($z>1$). We would produce a high-redshift counterpart to local large surveys of rich clusters like the well-studied Abell catalogue \citep{abe89}. We aim to address the following questions in detail: 1) How do galaxies and clusters co-evolve at early times? 2) How does environment (especially in overdense regions) affect star formation, enrichment, outflows and feedback processes? and 3) What is the time evolution of each of these processes? When and where do they peak? 

These are fundamental questions to understand the early stages of galaxy formation, an epoch that is still unknown, and that cannot be studied with current FIR/mm observatories in detail. This is only possible with a  sample of several thousand (proto)galaxy clusters from $z=1-10$ and a complete coverage of the spatial distribution of their individual members extending to several 10s of Mpc. In order to increase the number of spectroscopically confirmed galaxy (proto)clusters a fast survey sub-mm telescope of at least 40~m is needed. Such a telescope sizes guarantees the unambiguous identification of its individual members due to the relatively high angular resolution. To get spectroscopic redshifts of several hundred to thousand members per cluster, a multiplex instrument with up to several thousand elements per field of view is indispensable. First option would be a heterodyne instrument with a wide field of view and extremely large spectral bandwidth, however the costs would be exorbitant, so this will not be feasible. Thus, we will opt for the MKID bolometer technology which should provide integral field unit spectroscopic capabilities.

Both to guarantee spectroscopic redshifts from $z=1-10$ and a complete study of the most prominent lines emitted from the cold ISM such as multiple CO, the two [CI], the [CII], H$_{2}$O and HCN lines, the spectrometer should have a unprecedented bandwidth from 70 to 700~GHz. E.g., the brightest expected line emitted in the far-infrared is [CII] at 158$\mu$m. An instrument that had spectral coverage from 180 to 345~GHz could thus follow the early stages of cluster evolution from $z=4.5-10$, extending to higher frequencies ($\sim$700~GHz) would even allow us to map the peak of the star-formation and black hole activity of the Universe at $z=2$ \citep{mad14} with this line. Furthermore, such a set-up guarantees the detection of several CO lines and the so-called CO SLED (spectral line energy distribution) can be established \citep[e.g.,][]{wei07,dan09,dad15,ren19}. This enables us to securely determine physical properties such as the gas density, excitation temperature and even molecular gas mass. In addition, with both [CI] lines the total cold gas mass can be measured as well \citep{pap04,tom14}. Getting a complete picture of the cold ISM supported by a large sample size is indispensable to study in a statistical way if environment plays a role by measuring parameters such as star-formation efficiency, molecular gas fraction and excitation.

\begin{figure*}
\begin{centering} 
\includegraphics[width=\linewidth]{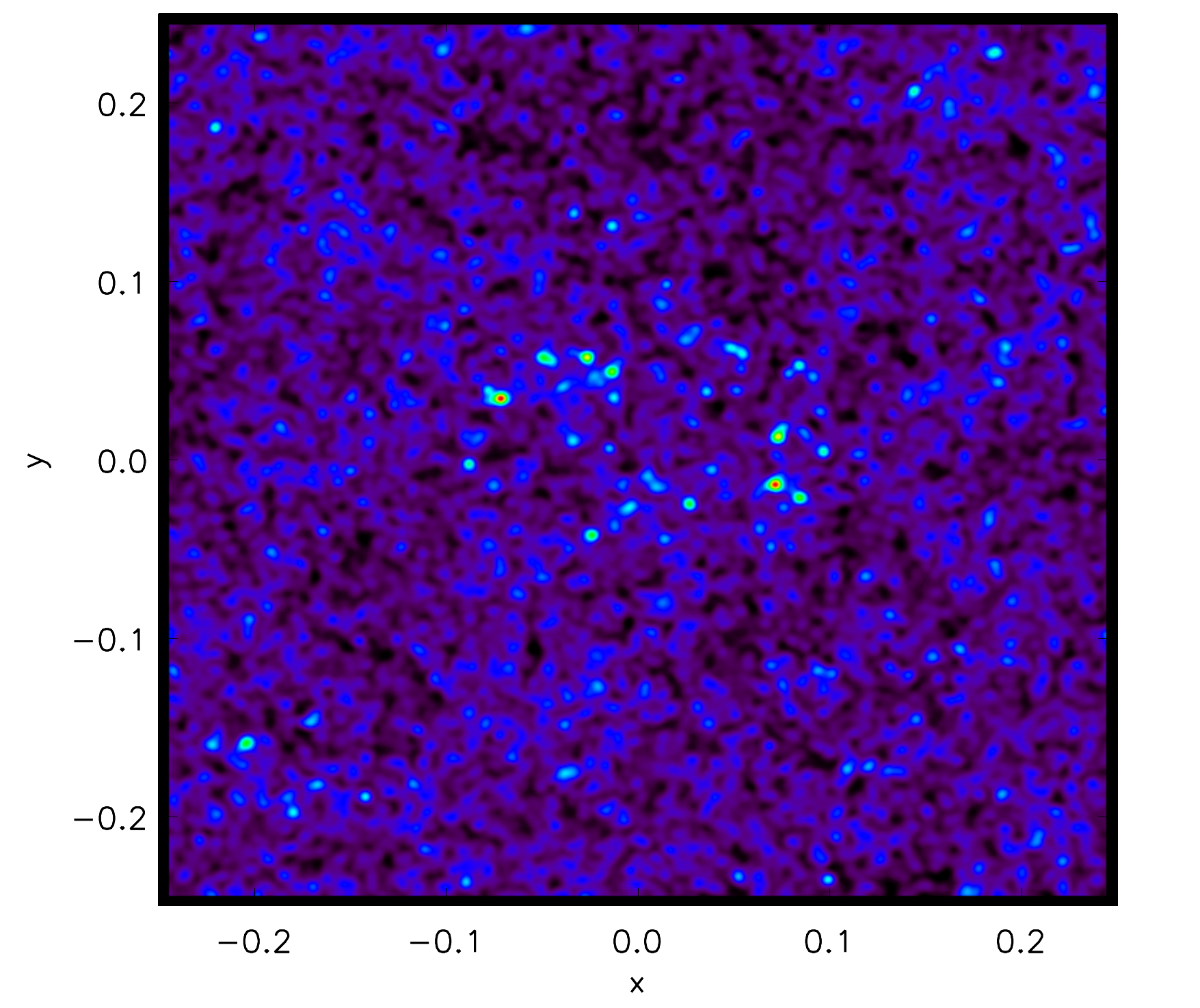}
\caption{Example mock image at 3~mm of the CO(3-2) flux of galaxies in a simulated $z=2$ protocluster, including homogeneous background noise. This $0.5\times0.5$ degree image constitutes a single pointing with a single-dish 50~m sub-mm telescope} \label{fig:protoclustersim}
\end{centering} 
\end{figure*}

Therefore, to make a big leap forward in understanding the evolution and formation of the largest structures and galaxies, a sub-mm telescope optimized for surveys is needed, i.e. a multiplexing instrument on a telescope with a single dish of at least 40~m. To visualize what such a telescope can achieve, a mock CO(3-2) image of a simulated protocluster at $z=2$ is shown in Fig.~\ref{fig:protoclustersim} (homogeneous noise is added), observed in a single pointing at 3~mm where the angular resolution would be of order of 15~arcsec for such a telescope, up to 7$\times$ better at higher frequencies. This mock image is derived from a light-cone constructed out of a semi-analytical galaxy formation model \citep{evk05} in which a cluster simulation (using the same model set-up) was inserted at $z=2$ (a few hundred cluster galaxies were added in this way, of which around 50 have sufficient signal-to-noise to be detectable at the depth of this particular mock image). ALMA would need several thousand hours to acquire such an image \citep{pop16}, whereas obtaining a large sample of these would require that to be a few hours at most.

We will target both already confirmed galaxy clusters (with known spectroscopic redshifts) and candidate clusters. In addition, we expect to discover new galaxy clusters, especially dust obscured ones with large molecular gas reservoirs contained in their individual members. At the time of conducting this survey with a large single dish telescope, a sample of several to a few ten thousand (proto)galaxy clusters from $z=1-10$ will exist coming from future surveys and missions such as LSST and Euclid, the latter one with a dedicated study to discover galaxy clusters at high redshift \citep{lau11}. Presently we already have a few thousand known candidate (proto)galaxy clusters from Planck \citep{pla15,pla16,gre18} and Hyper Suprime-Cam \citep{hig18,tos18}.

\section{Technical Requirements}
None of the current nor planned instruments within the next decade will fulfill the scientific requirements outlined in this manuscript for such an ambitious survey of galaxy clusters in the early Universe. We need a fast enough survey telescope that not only allows the study of confirmed galaxy clusters (with known spectroscopic redshifts) and candidate clusters but in addition will yield an unbiased survey of a significant area of the sky within a reasonable time span. Therefore, to conduct the survey and achieve the science goals following technical requirements should be fulfilled:

\begin{itemize}
\item a bolometer based on the MKID technology should have multiplex --- many thousand elements per field of view --- spectroscopic properties (similar to multi-object spectroscopy in the optical and near-infrared), unseen before in the (sub)millimeter window,
\item a large bandwidth from 70 to 700~GHz is indispensable for obtaining spectroscopic redshifts and study the physical properties of galaxy clusters at unprecedented detail simultaneously,
\item a field-of-view of $\sim$1~square degree to cover the typical size of  (proto)galaxy clusters in case of targeted observations,
\item a sub-mm telescope of at least 40~m is required to obtain individual detections of galaxy cluster members over the proposed frequency coverage, 
\item a spectral resolution of $500-1000$~km/s is needed to detect cluster galaxies and determine their redshifts (preferably from two or more lines), whereas $50-100$~km/s bins would be needed to resolve the lines and derived physical properties from these,
\item a survey speed of at least 15~arcmin$^2$ per minute is needed to be able to obtain a sample of several thousand galaxy clusters. 
\end{itemize}

\section{Concluding Remarks}
We make the case for a large aperture single-dish telescope such as the proposed Atacama Large-Aperture sub-mm/mm Telescope (AtLAST) to perform a large survey of distant galaxy clusters, a high-redshift counterpart to local large surveys of rich clusters like the well-studied Abell catalogue. Principal goals of such a survey are the evolution of clusters over cosmic time and the impact of environment on the evolution and formation of galaxies. Such a sub-mm facility with large-bandwidth multiplexing instrumentation will allow us to map and study a large, statistically significant and unbiased sample of thousands of distant galaxy clusters, each requiring a single pointing only to derive molecular gas masses and spectroscopic redshifts for their galaxy population. 

\pagebreak
{}


\begin{thebibliography}{}

\bibitem[Abell, Corwin \& Olowin(1989)]{abe89} Abell, G.~O., Corwin, H.~G., Olowin, O.~P.\ 1989, ApJS, 70, 1

\bibitem[Carilli \& Walter(2013)]{car13} Carilli, C.~L., \& Walter, F.\ 2013, ARA\&A, 51, 105 

\bibitem[Casey et al.(2014)]{cas14} Casey, C.~M., Narayanan, D., \& Cooray, A.\ 2014, PhR, 541, 45 

\bibitem[Casey(2016)]{cas16} Casey, C.~M.\ 2016, ApJ, 824, 36 

\bibitem[Coogan et al.(2018)]{coo18} Coogan, R.~T., Daddi, E., Sargent, M.~T., et al.\ 2018, MNRAS, 479, 703 

\bibitem[Daddi et al.(2015)]{dad15} Daddi, E., Dannerbauer, H., Liu, D., et al.\ 2015, A\&A, 577, A46 

\bibitem[Dannerbauer et al.(2009)]{dan09} Dannerbauer, H., Daddi, E., Riechers, D.~A., et al.\ 2009, ApJL, 698, L178 

\bibitem[Dannerbauer et al.(2014)]{dan14} Dannerbauer, H., Kurk, J.~D., De Breuck, C., et al.\ 2014, A\&A, 570, A55 

\bibitem[Dannerbauer et al.(2017)]{dan17} Dannerbauer, H., Lehnert, M.~D., Emonts, B., et al.\ 2017, A\&A, 608, A48 

\bibitem[Greenslade et al.(2018)]{gre18} Greenslade, J., Clements, D.L., Cheng, T., et al.\ 2018, MNRAS, 476, 3336

\bibitem[Hayashi et al.(2017)]{hay17} Hayashi, M., Kodama, T., Kohno, K., et al.\ 2017, APJL, 841, L21 

\bibitem[Hayashi et al.(2018)]{hay18} Hayashi, M., Tadaki, K.-i., Kodama, T., et al.\ 2018, ApJ, 856, 118 

\bibitem[Higuchi et al.(2018)]{hig18} Higuchi, R., Ouchi, M., Ono, Y., et al.\ 2018, ApJ, submitted (astro-ph/1801.00531) 

\bibitem[Ivison et al.(2013)]{ivi13} Ivison, R.~J., Swinbank, A.~M., Smail, I., et al.\ 2013, ApJ, 772, 137 

\bibitem[Kravtsov \& Borgani(2012)]{kra12} Kravtsov, A.~V., \& Borgani, S.\ 2012, ARAA, 50, 353

\bibitem[Laureijs et al.(2011)]{lau11} Laureijs, R., Amiaux, J., Arduini, S., et al.\ 2011, Euclid Definition Study Report, astro-ph/1110.3193 

\bibitem[Lutz et al.(2001)]{lut01} Lutz, D., Dunlop, J.~S., Almaini, O., et al.\ 2001, A\&A, 378, 70 

\bibitem[Madau \& Dickinson(2014)]{mad14} Madau, P., \& Dickinson, M.\ 2014, ARA\&A, 52, 415 

\bibitem[Muldrew et al.(2015)]{mul15} Muldrew, S.~I., Hatch, N.~A., \& Cooke, E.~A.\ 2015, MNRAS, 452, 2528 

\bibitem[Noble et al.(2017)]{nob17} Nobel, A.G., McDonald M., Muzzin A., et al.\ 2017, ApJL, 842, 2

\bibitem[Noble et al.(2019)]{nob19} Noble, A.~G., Muzzin, A., McDonald, M., et al.\ 2019, ApJ, 870, 56 

\bibitem[Overzier(2016)]{ove16} Overzier, R.~A.\ 2016, ARA\&A, 24, 14 

\bibitem[Papadopoulos et al.(2004)]{pap04} Papadopoulos, P.~P., Thi, W.-F., \& Viti, S.\ 2004, MNRAS, 351, 147 

\bibitem[Planck Collaboration et al.(2015)]{pla15} Planck Collaboration, Aghanim, N., Altieri, B., et al.\ 2015, A\&A, 582, A30 

\bibitem[Planck Collaboration et al.(2016)]{pla16} Planck Collaboration, Ade, P.~A.~R., Aghanim, N., et al.\ 2016, A\&A, 596, A100

\bibitem[Popping et al.(2016)]{pop16} Popping, G., van Kampen E., Decarli, R., et al.\ 2016, MNRAS, 461, 99

\bibitem[Renaud et al.(2019)]{ren19} Renaud, F., Bournaud, F., Daddi, E., \& Wei{\ss}, A.\ 2019, A\&A, 621, A104 

\bibitem[Rudnick et al.(2017)]{rud17} Rudnick, G., Hodge, J., Walter, F., et al.\ 2017, ApJ, 849, 27 

\bibitem[Stach et al.(2017)]{sta17} Stach, S.~M., Swinbank, A.~M., Smail, I., et al.\ 2017, ApJ, 849, 154 

\bibitem[Tacconi et al.(2018)]{tac18} Tacconi, L.~J., Genzel, R., Saintonge, A., et al.\ 2018, ApJ, 853, 179 

\bibitem[Tadaki et al.(2019)]{tad19} Tadaki, K.-I., Kodama, T., Hayashi, M., et al.\ 2019, PASJ, in press (astro-ph/1901.07173)

\bibitem[Tomassetti et al.(2014)]{tom14} Tomassetti, M., Porciani, C., Romano-D{\'{\i}}az, E., Ludlow, A.~D., \& Papadopoulos, P.~P.\ 2014, MNRAS, 445, L124 

\bibitem[Toshikawa et al.(2018)]{tos18} Toshikawa, J., Uchiyama, H., Kashikawa, N., et al.\ 2018, PASJ, 70, S12 

\bibitem[van Kampen et al.(2005)]{evk05} van Kampen, E., Percival W.J., Crawford M., et al.\ 2005, MNRAS, 359, 469

\bibitem[Weiss et al.(2007)]{wei07} Weiss, A., Downes, D., Walter, F., \& Henkel, C.\ 2007, From Z-Machines to ALMA: (Sub)Millimeter Spectroscopy of Galaxies, 375, 25 


\end{thebibliography}
\end{document}